\documentclass[reprint,prx,amsmath,amssymb,superscriptaddress]{revtex4-2}

\usepackage{array}
\usepackage{multirow}
\usepackage{amsmath}
\usepackage{amsfonts}
\usepackage{graphicx}
\usepackage{xcolor}
\usepackage[version=4]{mhchem}
\usepackage{bm}
\usepackage{bbm}
\usepackage{physics}
\usepackage{subfigure}
\usepackage{float}
\usepackage{hyperref}
\hypersetup{
	colorlinks=true,
	linkcolor=blue,
	filecolor=gray,
	urlcolor=blue,
	citecolor=blue,
}

\begin{document}
\title{Chiral channel network from magnetization textures in 2D \ce{MnBi_2Te_4}}
\author{Chengxin Xiao}
\affiliation{Department of Physics, The University of Hong Kong, Hong Kong, China }
\affiliation{HKU-UCAS Joint Institute of Theoretical and Computational Physics at Hong Kong, China }

\author{Jianju Tang}
\affiliation{Department of Physics, The University of Hong Kong, Hong Kong, China }
\affiliation{HKU-UCAS Joint Institute of Theoretical and Computational Physics at Hong Kong, China }

\author{Pei Zhao}
\affiliation{Department of Physics, The University of Hong Kong, Hong Kong, China }
\affiliation{HKU-UCAS Joint Institute of Theoretical and Computational Physics at Hong Kong, China }

\author{Qingjun Tong}
\affiliation{School of Physics and Electronics, Hunan University, Changsha 410082, China}

\author{Wang Yao}
\affiliation{Department of Physics, The University of Hong Kong, Hong Kong, China }
\affiliation{HKU-UCAS Joint Institute of Theoretical and Computational Physics at Hong Kong, China }

\begin{abstract}
    When atomically thin van der Waals (vdW) magnet forms long-period moir\'e pattern with a magnetic substrate, the sensitive dependence of interlayer magnetic coupling on the atomic registries can lead to moir\'e defined magnetization textures in the two-dimensional (2D) magnets. The recent discovery of 2D magnetic topological insulators such as \ce{MnBi_2Te_4} leads to the interesting possibility to explore the interplay of such magnetization textures with the topological surface states, which we explore here with a minimal model established for 2D \ce{MnBi_2Te_4}. The sign flip of the exchange gap across a magnetization domain wall gives rise to a single in-gap chiral channel on each surface. In the periodic magnetization textures, such chiral spin channels at the domain walls couple to form a network and superlattice minibands emerge. We find that in magnetization textures with closed domain wall geometries, the formed superlattice miniband is a gapped Dirac cone featuring orbital magnetization from the current circulation in the close loops of chiral channels, while in magnetization textures with open domain wall geometries, gapless mini-Dirac cone is found instead. The miniband Bloch states feature a spatial texture of spin and local current density, which are clear manifestation of the spin-momentum locked chiral channels at the domain walls. The results suggest a new platform to engineer spin and current flows through the manipulation of magnetization domains for spintronic devices.
\end{abstract}
\maketitle

\section{Introduction}

The study of band topology and magnetic order in condensed matter systems are of great interest for their potential applications in low-power-consumption electronics and spintronics \cite{fan2016spintronics,pesin2012spintronics,vzutic2004spintronics}. Combining these two ingredients in a single quantum system will open up a new realm to explore exotic quantum state of matter, including quantum anomalous Hall (QAH) effects with topologically protected chiral edge states \cite{haldane1988model,chang2013experimental}, topological axion state with quantized magnetoelectric effects \cite{qi2008topological,essin2009magnetoelectric,mong2010antiferromagnetic}, and topological superconductor with Majorana fermions obeying non-abelian statistics \cite{alicea2012new}. A recent advance along this direction is the discovery of intrinsic magnetic topological insulators in vdW layered material \ce{MnBi_2Te_4}, which shows 2D ferromagnetism within individual layer and antiferromagnetic (AFM) interlayer coupling \cite{lee2013crystal,zhang2019topological,gong2019experimental,otrokov2019prediction,ge2019high,yan2019crystal,hao2019gapless}. The band topology gives rise to surface Dirac cones, which are gapped by the exchange interaction with the magnetic order. Inside this exchange gap the Dirac cone has a half-quantized topological charge with sign controlled by the out-of-plane magnetization orientation. With the layered AFM order, the topology of the surface states becomes dependent on the thickness. With the magnetization in top and bottom layer having opposite (same) sign in even (odd) numbered layers, the gapped Dirac cones at top and bottom surfaces either add or cancel in their contributions to the Hall conductivity. This intriguing property has been utilized to engineer axion insulator state and QAH state, as observed in recent experiments \cite{liu2020robust,deng2019magnetic}. The competition of the interlayer AFM coupling and the Zeeman energy in external magnetic field also makes possible different topological phases associated with the various layered magnetic orders \cite{li2019magnetically,liu2020helical}.

The coupling with magnetic order further makes possible engineering of topological electronics state on the surface through manipulation of the in-plane magnetization textures such as domain walls.
Another unique opportunity that can be explored in the vdW layered magnetic topological insulators is the formation of long-period moir\'e pattern due to small twisting and lattice mismatch between the layers or with the substrate\cite{geim2013van,ponomarenko2013cloning,dean2013hofstadter,hunt2013massive,gorbachev2014detecting,zhang2017interlayer}.
In such moir\'e pattern, the interlayer atomic registry in any local region resembles that of lattice-matched stacking, whereas the stacking order changes smoothly over long distance with a periodicity that can range from a few nm to tens of nm \cite{tong2017topological,yu2017moire}.
The interlayer magnetic coupling can have sensitive dependence on the stacking registry \cite{jiang2019stacking,soriano2019interplay,sivadas2018stacking,wang2018very,klein2019enhancement,sun2019giant,song2019switching,li2019pressure,chen2019direct}.
As a result, the magnitude and sign of interlayer magnetic coupling get spatially modulated in the moir\'e pattern, which can lead to lateral magnetization textures including the magnetic bubble (MB) lattice and the skyrmion (SK) lattice with the moir\'e periodicity \cite{tong2018skyrmions}.
Such magnetization textures defined by the moir\'e, combined with the dramatic tunability of the latter by hetero-strain, point to exciting possibilities to tailor the surface electronic states in the atomically thin vdW magnetic topological insulators.

In this work, we study the surface Dirac cone under the exchange coupling with periodic magnetization textures. The sign flip of the exchange gap across a magnetization domain wall leads to a quantized change of the topological charge, giving rise to a single chiral channel inside the exchange gap on each surface. Electrons propagate unidirectionally along the domain wall, with spin locked to the perpendicular direction in plane. In the periodic magnetization textures, such chiral spin channels at the domain walls form a network and superlattice minibands emerge out of their coupling. We find that in magnetization textures with closed domain wall geometries, e.g. the SK lattice, the formed superlattice miniband is a gapped Dirac cone featuring orbital magnetization from the current circulation in the close loops of chiral channels. In magnetization textures with open domain wall geometries, e.g. the MB lattice, gapless mini-Dirac cone is found instead.
The miniband Bloch states feature a spatial texture of spin and local current density, where the spin-momentum locked chiral channels pinned to the domain walls are clearly manifested.
These results suggest a new platform to engineer spin and current flows through the manipulation of magnetization domains for spintronic devices.
A promising system to explore these interesting interplay between topological surface state and the magnetization texture is bilayer \ce{MnBi_2Te_4}, where the lateral magnetization textures can be generated by moir\'e pattern formed between the bilayer and a suitable magnetic substrate.

The rest of the paper is organized as follows. In Sec.~\ref{structure}, we give a brief account of the magnetic configurations and topological surface states in 2D \ce{MnBi_2Te_4}, and two typical magnetization textures that can arise from the laterally modulated interlayer magnetic coupling in a moir\'e pattern. The property of in-gap chiral state at a single magnetization domain wall is also discussed. Sec.~\ref{miniband} investigates the minibands formed when the surface Dirac cone is subjected to the periodically modulated exchange field from the magnetization textures. The local characterization of wavefunctions shows how the miniband dispersion and orbital magnetization arise from the chiral channels at the domain walls in different magnetization textures. Finally, a symmetry analysis on the miniband spectrum and a summary are given in Sec.~\ref{summary}.

\section{Surface Dirac cone, chiral domain wall state, and moir\'e defined magnetization textures}\label{structure}

\begin{figure}[htbp]
    \centering
    \includegraphics[width=\linewidth]{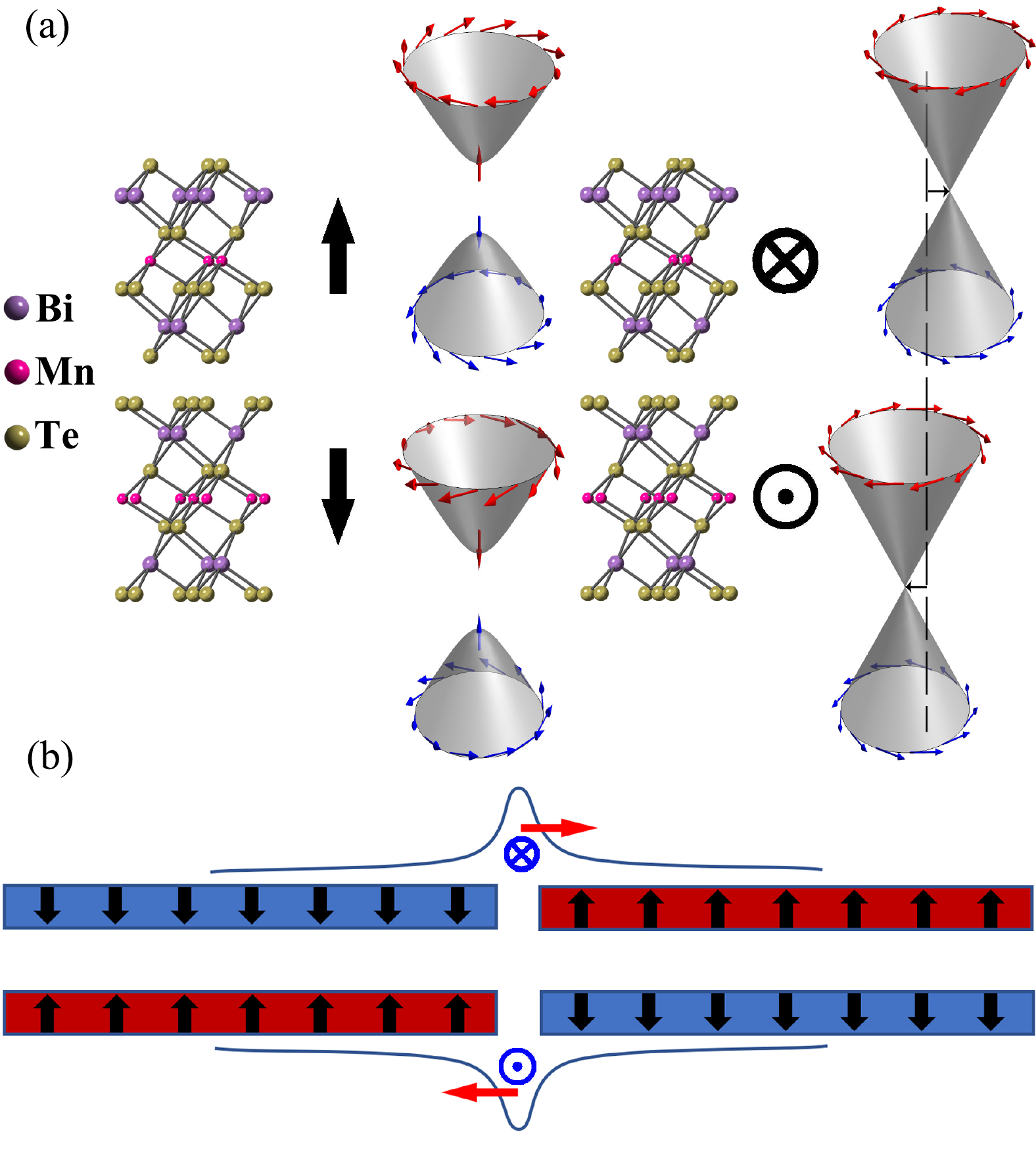}
    \caption{(a) Schematic of \ce{MnBi_2Te_4} bilayer and its surface state spectrum in AFM-z (left) and AFM-x (right) order. The black arrows and circles with cross and dot representing inward and outward describe the surface magnetization. The spectrum is gapped for AFM-z and gapless for AFM-x order. The spins (red and blue arrows) in the surface state are locked with their momentum. (b) Schematic of a magnetic domain wall in \ce{MnBi_2Te_4} bilayer and the corresponding topological domain wall states on top and bottom surfaces, with the spin (red arrows) locked to the velocity (blue circles with cross and dot representing inward and outward).}\label{MBT}\label{domain_wall}
\end{figure}

\ce{MnBi_2Te_4} is a vdW material formed by ABC-stacked Te-Bi-Te-Mn-Te-Bi-Te septuple layers in a rhombohedral structure with the space group R\(\bar 3\)m~\cite{lee2013crystal}, as shown in Fig.~\hyperref[MBT]{1(a)}. The in-plane triangular lattice constant is $a = 4.334$\AA~\cite{lian2020flat}. Theoretical predictions and experimental observations point to a simple A-type AFM ground state with an out-of-plane ferromagnetic (FM) order within the layer, and AFM order between the layers.

In the (111) surface, there exists a surface Dirac cone with its gap depending on the orientation of the surface magnetization. For an out-of-plane magnetization, first-principle calculations gives a gap of \(\Delta \approx 50\) meV \cite{gong2019experimental}. While with in-plane magnetization, e.g. aligned to an in-plane magnetic field, the surface spectrum is a translated gapless Dirac cone \cite{hao2019gapless} (Fig. \hyperref[domain_wall]{1(a)}). Because the two surfaces have the same (opposite) magnetization for an even (odd) layer thin film, we focus on a single-surface in the following. The low-energy effective Hamiltonian for a single surface can be written as,

\begin{equation}
    \mathcal{H}_S=\hbar v (\sigma_x k_y - \sigma_y k_x)+ g\bm{M}\cdot\bm{\sigma}
    \label{surface}
\end{equation}

\noindent where \(\bm{\sigma}\) is the Pauli matrix for spin degree of freedom, \(\hbar v\approx 2\text{eV}\cdot\text{\AA}\) is the Fermi velocity, and the last term is the exchange coupling to the surface magnetization $\bm{M}$, with $g$ being the coupling constant. One can see that the in-plane components of $\bm{M}$ shift the Dirac point in the momentum space, while the out-of-plane component creates a gap.

The gapped Dirac surface state can be characterized by a topological number $C_n=\frac{1}{2\pi}\int\Omega_n d\bm{k}$, where the Berry curvature is defined by $\Omega_n=\hat{z}\cdot\bm{\nabla}_{\bm{k}}\times\left\langle u_{n}(\bm{k})\middle| i\bm{\nabla}_{\bm{k}}\middle| u_{n}(\bm{k})\right\rangle$ and $u_{n}(\bm{k})$ is the eigenfunction of Eq. \eqref{surface} with band index $n$. For the Dirac cone described by Eq. \eqref{surface}, the Berry curvature is $\Omega_v(\bm{M})=\hbar ^2 v^2gM_z/(2\varepsilon^3)$ for the valence band, where $\varepsilon=\sqrt{\hbar ^2 v^2(\bm{k}-\bm{k}_0)^2+g^2M_z^2}$ and $\bm{k}_0=(gM_y/\hbar v,-gM_x/\hbar v)$. The integration of the Berry curvature gives a topological number $C_v=\frac{1}{2}sgn(M_z)$, with a sign determined by the out-of-plane component of the magnetization $\bm{M}$.

For a surface with a magnetic domain wall, the topological charge has a quantized difference across the domain wall: $Q=C_v\left(M_z^+\right)-C_v\left(M_z^-\right)=\pm1$. The bulk-edge correspondence then dictates an in-gap chiral state along the domain wall.
This is similar in origin to the well-studied kink states in bilayer graphene~\cite{martin2008topological,jung2011valley,zhang2013valley,li2014spontaneous,vaezi2013topological,ju2015topological,li2016gate,yin2016direct,yao2009edge}.
The kink states in graphene comes in pairs with opposite chirality at the two valleys, and also have a spin degeneracy, so back scattering is still allowed. In contrast, 2D \ce{MnBi_2Te_4} also has a pair of surface Dirac cones, but spatially separated to the top and bottom surfaces respectively. At each surface, the magnetization domain wall localizes a single chiral channel completely free from back scattering, which is highly advantageous as a conduction channel or spin channel. Fig.~\hyperref[domain_wall]{1(b)} schematically shows an AFM domain wall in a bilayer \ce{MnBi_2Te_4}, which localizes one chiral channel on the top and bottom surfaces respectively with opposite chirality.

We note that such AFM domain wall is recently observed in bilayer \ce{CrI_3} which has the same layered magnetic order, i.e.\ out-of-plane ferromagnetism in individual layer and interlayer AFM coupling. The domain wall can form due to imperfections in fabrication. Experiments in top and bottom gated bilayer \ce{CrI_3} have also shown that gate voltage can reversibly switch between the different AFM configurations~\cite{song2019voltage},
as well as between the AFM and ferromagnetic layered configurations near the critical magnetic field~\cite{jiang2018electric}.
It is therefore possible to realize configurable AFM domain wall or AFM/FM domain wall on demand using split double gate design.

For a magnetization domain wall parallel to the y-axis, one can solve from Eq. \eqref{surface} the wavefunction of the chiral state,

\begin{equation}
    \varphi_{k_y}(x,y) \propto e^{ik_y y}\left(\begin{array}{c}
            {1} \\
            {\eta}
        \end{array}\right) \exp \left[-\eta\int_0^{x} \frac{gM_z(x^{\prime})}{\hbar v} d x^{\prime}\right]
\end{equation}

\noindent with the dispersion $E\left(k_y\right)=\eta \hbar vk_y$ and $\eta=sgn\left(M_z^+\right)$ determined by the sign of the gap term at the right-hand side of domain wall. The chiral mode moves with a velocity $\eta v$ along y-direction. The spin expectation value of the domain wall state is $\bm{s}=\eta\hat{x}$, whose direction is perpendicular to the velocity direction, as a manifestation of the spin-momentum locking in the parent Dirac cone. For a sharp domain wall, the full width at half hight of this chiral state is about \(2\log 2\hbar v/g\), which is  $\approx 10\text{nm}$ for a \(g= 25\text{meV}\).

\begin{figure}[htbp]
    \includegraphics[width=\linewidth]{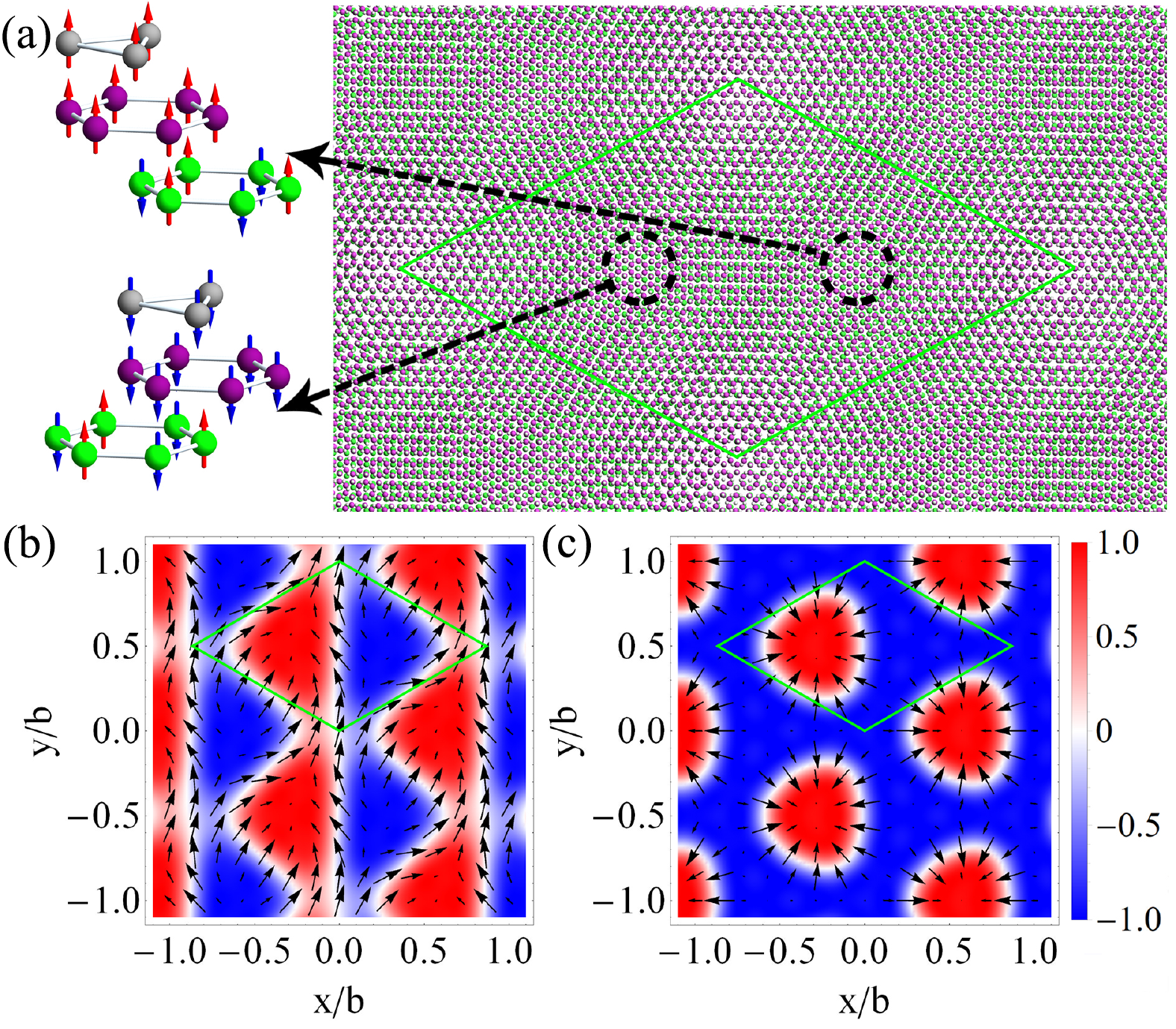}
    \caption{(a) Schematic of a moir\'e pattern formed by an \ce{MnBi_2Te_4} bilayer stacked on an FM medium layer \ce{CrI_3} twisted on an AFM substrate \ce{MnPTe_3}. The green diamond denotes a moir\'e unit cell with periodicity $b$. Only one layer of Mn atoms from \ce{MnBi_2Te_4} are shown as the grey spheres, and the magnetic atoms on the FM medium layer and AFM substrate are denoted by the purple and green spheres. At the two local regions marked by the dashed circles, the \ce{Mn} atoms in \ce{MnBi_2Te_4} and \ce{Cr} atoms in \ce{CrI_3} sit on top of the spin up and spin down sublattices of the AFM substrate respectively. Consequently, the local ferromagnetic order in the \ce{MnBi_2Te_4} layer tends to be pinned in the up and down directions respectively at these two local regions by the proximity magnetic coupling from the substrate. (b-c) The competition of this proximity field with the in-plane exchange and anisotropy magnetic interaction can turn the uniform magnetization order in the \ce{MnBi_2Te_4} bilayer into either an MB lattice (b), or a SK lattice (c) (see \hyperref[B]{Appendix B} for detailed calculations of the magnetization textures). The z-component of the magnetization is color coded, and the in-plane component is denoted by the arrows near the domain walls, and the two examples of textures are reproduced using parameters from \cite{tong2018skyrmions}.}\label{mass}
\end{figure}

Motivated by the emergence of the chiral channel at the domain wall between spin up and down magnetization, we now turn to the periodic magnetization textures which can be formed in a long-period moir\'e pattern between the bilayer \ce{MnBi_2Te_4} and a magnetic substrate.
In \ce{MnBi_2Te_4}, because the Mn atoms are sandwiched by six layers of Bi and Te atoms, a direct magnetization reversal via proximity effect from an antiferromagnetic substrate is difficult. However, a recent work shows that the \ce{MnBi_2Te_4}/\ce{CrI_3} heterostructure has a strong interlayer FM coupling and this coupling is nearly stacking independent \cite{fu2020exchange}. We therefore resort to \ce{CrI_3} as a medium layer to generate magnetization texture in \ce{MnBi_2Te_4}. In an \ce{MnBi_2Te_4}/\ce{CrI_3} heterostructure, when magnetization is reversed in \ce{CrI_3}, the magnetization is also reversed in \ce{MnBi_2Te_4} because of the strong interlayer coupling, as schematically shown in Fig. \hyperref[mass]{2(a)}. The magnetization texture in \ce{CrI_3} can be formed in twisted bilayer \ce{CrI_3} or \ce{CrI_3}/\ce{MnPTe_3} heterostructure. In the former, the rhombohedral stacking favors interlayer FM coupling and the monoclinic stacking favors AFM coupling \cite{jiang2019stacking,soriano2019interplay,sivadas2018stacking,wang2018very}. In the latter, when the magnetic atoms \ce{Cr} in \ce{CrI_3} sit on top of the Mn atoms in \ce{MnPTe_3} with opposite magnetic moments, the first-principles results show that the ferromagnetic order in \ce{CrI_3} tends to be pinned in the opposite directions when the magnetic order in the antiferromagnetic substrate is fixed (see \hyperref[A]{Appendix A}). When \ce{MnBi_2Te_4} is stacked on this texture, because of the strong interlayer FM coupling between \ce{MnBi_2Te_4} and \ce{CrI_3}, the magnetization texture would also be formed in \ce{MnBi_2Te_4}.

There are two types of magnetization textures distinct by different topological winding of the magnetization in a moir\'e unit cell, and different domain wall geometries. As shown in Fig.~\hyperref[mass]{2}, in the topologically trivial MB lattice, the in-plane magnetization around the domain wall tends to order in the same direction. There are two types of domain walls along y direction, one is straight and the other is of zigzag shape. Furthermore, the out-of-plane magnetization $M_z(\bm{r})$ in a moir\'e period is antisymmetric. The other phase is SK lattice, which is topologically nontrivial with the magnetization texture around the domain walls forming a vortex structure. This texture has $C_3$ rotation symmetry and the domain walls form close loops. In this case, the magnetization texture does not have the antisymmetry.

\section{Network of chiral domain wall states in the periodic magnetization textures}\label{miniband}

With a single domain wall introducing an in-gap chiral channel, the moir\'e-patterned magnetization textures with the periodic domains then give rise to a network of chiral channels which are coupled due to the finite size of the domains. In this section, we will examine the moir\'e minibands formed by this network of chiral channels, and the local characterization of wavefunctions, and show how the miniband dispersion and orbital magnetization arise in different magnetization textures.

The appearance of a magnetization texture breaks the original translation symmetry of the surface states of \ce{MnBi_2Te_4}, and introduces a spatially periodically varying Zeeman term $g\bm{M}(\bm{r})\cdot\bm{\sigma}$, where $\bm{r}$ is the position vector. We first look at the effect from the periodically modulated $M_z(\bm{r})$, as the topological property of the surface Dirac cone is determined by this component, and the moir\'e magnetization texture is also primarily along z-direction (Figs. \hyperref[mass]{2(b)} and \hyperref[mass]{(c)}). In-plane magnetization is relevant only at the domain walls, and its effect will be discussed in Sec.~\ref{summary}. In the case of MB magnetization texture, it can cause a displacement of the minibands in the momentum space. The Hamiltonian describing the surface state in a spatial texture of $M_z(\bm{r})$ is,
\begin{equation}
    \mathcal{H}_S= -i\hbar v (\sigma_x \partial_y - \sigma_y \partial_x) + \Delta_z(\bm{r})\sigma_z
    \label{H}
\end{equation}
where $\Delta_z(\bm{r}) = gM_z(\bm{r})$. The moir\'e periodicity allows us to expand \(\Delta_z(\bm r)=\sum _{\bm G} \tilde{\Delta}_z(\bm G) e^{i\bm G \cdot \bm r}\), where \(\bm{G}\) is the reciprocal lattice vector of the moir\'e superlattice.

The low-energy physics is obtained by diagonalizing the Hamiltonian in the plane wave basis,
\begin{equation}
    \begin{aligned}
        \mel**{\bm k+\bm G',s'}{\mathcal{H}_S}{\bm k +\bm G,s} & = \delta_{\bm{GG}'}\hbar v \left[\sigma_x^{s's} (k_y+G_y)\right. \\
        \left. - \sigma_y^{s's} (k_x+G_x)\right]               & + \tilde{\Delta}_z(\bm G'-\bm G) \sigma_z^{s's}
    \end{aligned}\label{matrix}
\end{equation}
\noindent where \(\ket{\bm k,s}=\frac{1}{\sqrt{NA}} e^{i\bm k \cdot \bm r}\psi_s\) with $\psi_{s=\uparrow,\downarrow}$ being the eigenvectors of $\sigma_z$. \(N\) is the number of supercells and \(A\) is the area of a supercell. The real-space wave function is constructed using the eigenvectors of the matrix,
\begin{equation}
    \psi_{n,\bm{k}}\left(\bm{r}\right)=\sum_{\bm{G}} e^{i\left(\bm{k}+\bm{G}\right)\cdot\bm{r}}\binom{\psi_{n,\bm{k},\bm{G},\uparrow}}{\psi_{n,\bm{k},\bm{G},\downarrow}}.
\end{equation}

To study the distribution of the domain wall states in the presence of the magnetization texture and its chirality and interplay with spins, in the following, we investigate the miniband states in terms of the local density of states (LDOS) \(\rho (\bm{r})=\abs{\psi(\bm r)}^2\), local spin expectation \(\bm{s} (\bm{r})=\psi(\bm r)^\dagger \bm{\sigma} \psi(\bm r)/\abs{\psi(\bm r)}^2\) and electric current density \(\bm j (\bm{r})=-e{\psi_{n,\bm{k}}\left(\bm{r}\right)}^\dagger\bm{v}\psi_{n,\bm{k}}\left(\bm{r}\right)=-ev[-\psi(\bm r)^\dagger \sigma_y \psi(\bm r)\hat{\bm x}+\psi(\bm r)^\dagger \sigma_x \psi(\bm r)\hat{\bm y}]\).

\subsection{MB magnetization texture}

\begin{figure}[htbp]
    \centering
    \includegraphics[width=\linewidth]{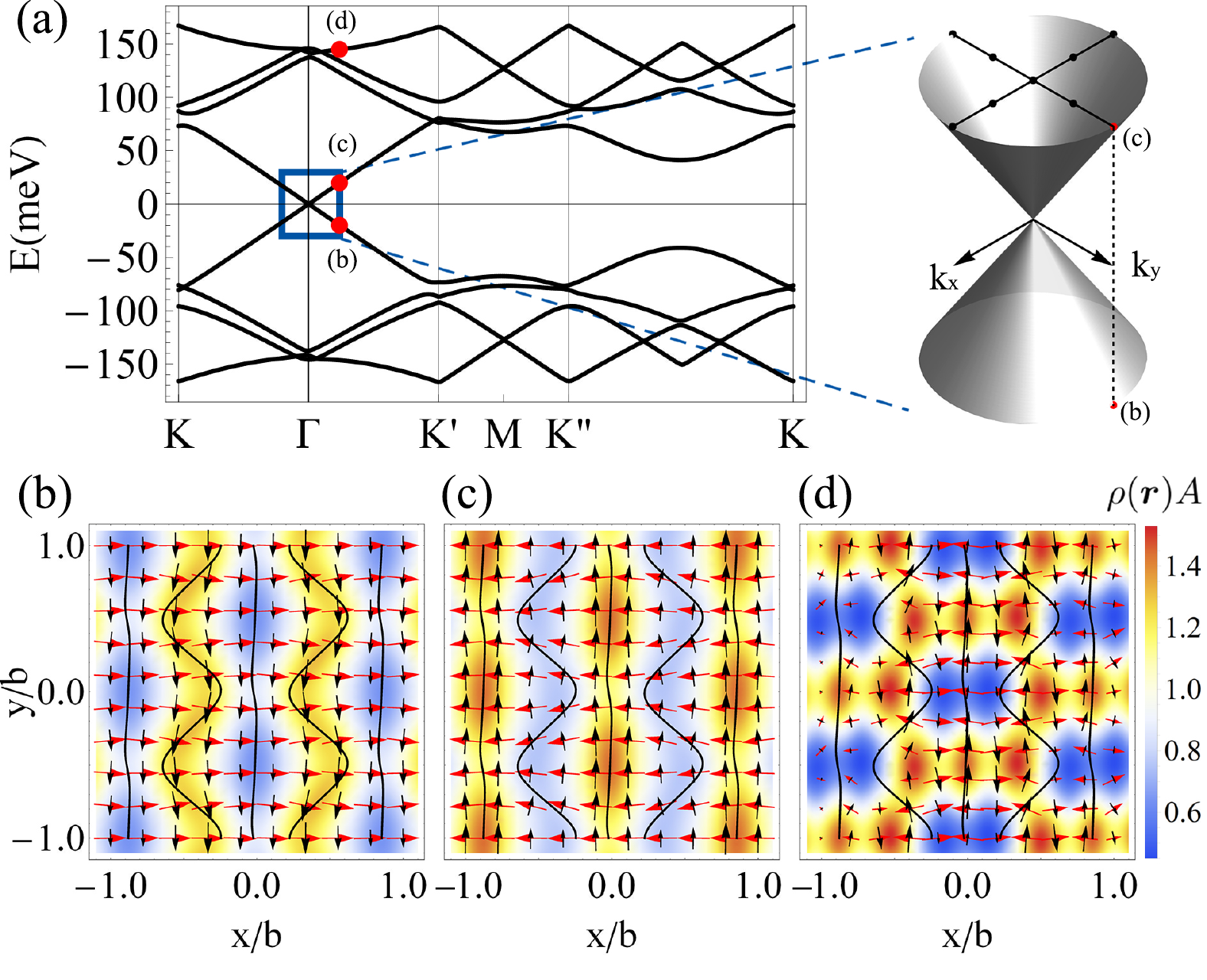}
    \caption{(a) The moir\'e minibands of the Dirac surface states subjected to the exchange field from an MB magnetization texture of moir\'e periodicity \(b=10\)nm. The right panel shows a full dispersion around the Dirac point. (b-d) LDOS \(\rho(\bm{r})\) (background), electric current density \(\bm j(\bm{r})\) (black arrows) and in-plane components of local spin expectation \(\bm s(\bm{r})\) (red arrows) with wave vector \(k=1/b\) in y direction. The corresponding wave vector and band index are indicated by the red dots in (a). The black lines are the domain walls. The parameters used are \(\hbar v = 2\text{eV}\cdot\text{\AA}\) and \(g=25\mathrm{meV}\).}\label{mb10}
\end{figure}

\begin{figure*}[htbp]
    \centering
    \includegraphics[width=0.9\linewidth]{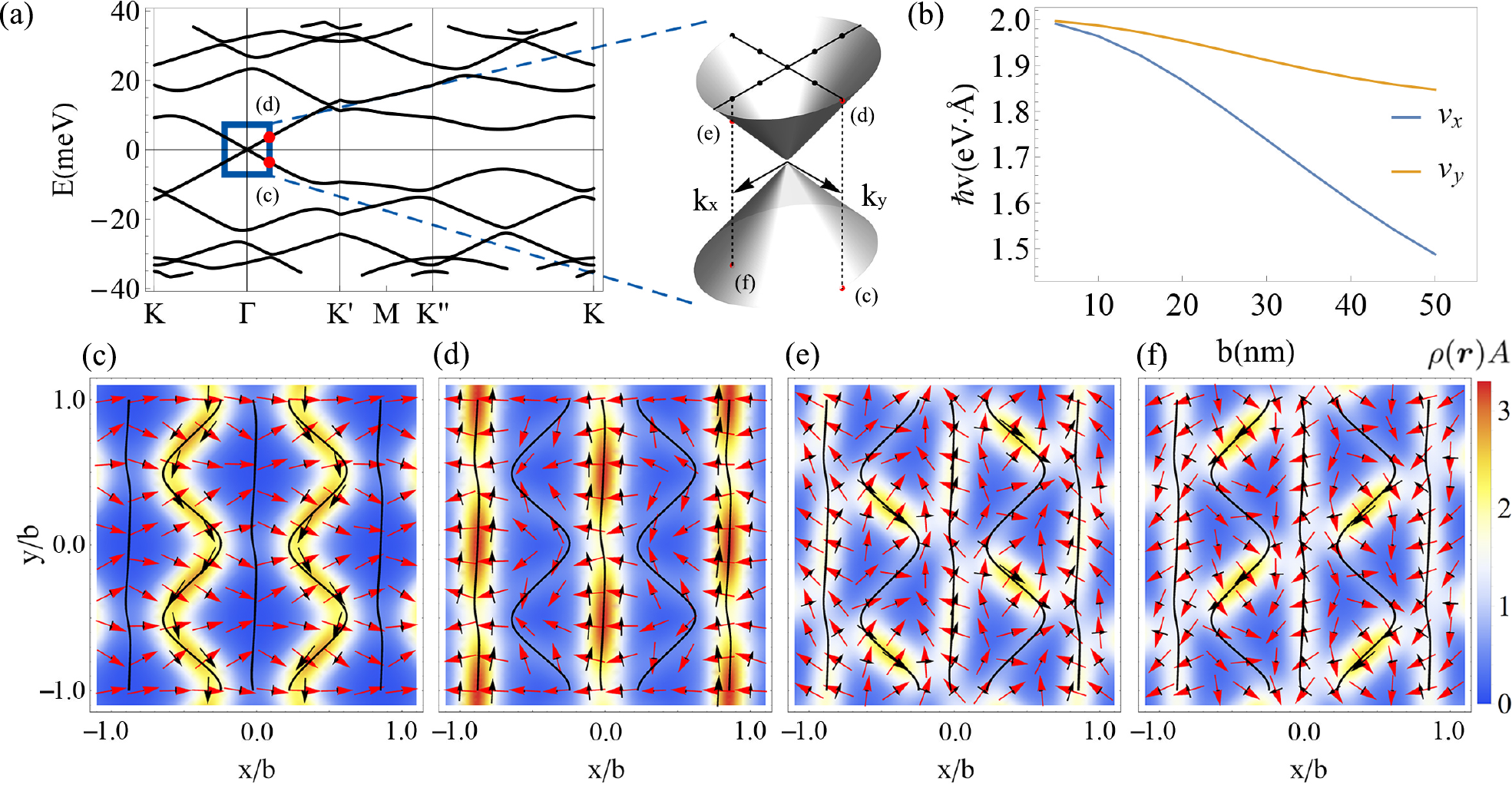}
    \caption{(a) The moir\'e minibands for an MB magnetization texture of moir\'e periodicity \(b=50\)nm. The right panel shows a full dispersion around the Dirac point. (b) The velocity in x and y direction near the Dirac point as a function of moir\'e periodicity. (c-f) LDOS \(\rho(\bm{r})\) (background), electric current density \(\bm j(\bm{r})\) (black arrows) and in-plane components of local spin expectation \(\bm s(\bm{r})\) (red arrows) with wave vector \(k=1/b\) in x and y direction. The corresponding wave vector and band index are indicated by the red dots in (a). The black lines are the domain walls. The parameters used are \(\hbar v = 2\text{eV}\cdot\text{\AA}\) and \(g=25\mathrm{meV}\).}\label{mb50}
\end{figure*}

The MB magnetization texture features straight and zigzag shape domain walls along y-direction, which shall localize an array of 1D chiral channels. The close proximity between the neighboring straight and zigzag domain walls at the corners of the moir\'e supercell (c.f. Fig. \hyperref[mass]{2(b)}) can couple the chiral channels, giving rise to a 2D dispersion. Figs.~\hyperref[mb10]{3(a)} and \hyperref[mb50]{4(a)} show the moir\'e minibands for Dirac surface state in MB magnetization texture of periodicity \(b=10\)nm and \(b=50\)nm respectively. Near zero energy, the miniband features a gapless Dirac point at $\mathrm{\Gamma}$ point of the mini-Brillouin zone, which is consistent with the antisymmetric $M_z(\bm{r})$ of the MB texture (detail of the symmetry analysis is given in Sec.~\ref{summary}).

At small periodicity, this Dirac cone in the miniband is nearly isotropic, although it originates from the chiral channels all along the y directions. The mini-Dirac cone has nearly the same Fermi velocity as the original surface Dirac cone as if the effect of the magnetization is averaged out. Figs. \hyperref[mb10]{3(b-d)} show the characters of several wavefunctions in terms of the LDOS \( \rho (\bm{r})\) (color map), in-plane component of local spin expectation \(\bm s (\bm{r}) \) (red arrows), and local current density \(\bm j (\bm{r})\) (black arrows). The two states near the Dirac point clearly have their origin from the in-gap chiral channels localized at the zigzag (Fig. \hyperref[mb10]{3(b)}) and the straight (Fig. \hyperref[mb10]{3(c)}) domain walls respectively, with opposite Dirac velocity. For this short-period magnetization texture, the width of the domain wall states (c.f. Sec.~\ref{structure}) is comparable with the periodicity. As a result, the wave function is largely delocalized and has significant distribution between the domain walls. In this case, the LDOS shows nearly straight stripes along y-direction at both the straight and zigzag domain walls, not much affected by the domain wall shape, and the local current density is also primarily aligned with the wavevector of the state (black arrows in Figs.~\hyperref[mb10]{3(b)} and \hyperref[mb10]{(c)}). Fig. \hyperref[mb10]{3(d)} is in a higher energy miniband, which does not have the chiral state origin, and the wavefunction is primarily distributed within the domains. The current density \(\bm j (\bm{r})\) is found always orthogonal to the in-plane component of the local spin \(\bm{s} (\bm{r})\), which can be expected from their expressions given earlier, as manifestation of the spin-momentum locking.

Anisotropy develops with the increase of the moir\'e periodicity. As shown in Figs.~\hyperref[mb50]{4(a)} and \hyperref[mb50]{(b)}, when the periodicity gets significantly larger than the width of the chiral state ($\sim 10$ nm), the Dirac cone is  anisotropic with the Fermi velocity in y direction significantly larger than the one in x direction. This is consistent with the fact that the coupling between the chiral channels are reduced as the domain walls get more spatially separated, recovering the 1D nature of the chiral channels. And for the states near the Dirac point, the LDOS shows that the wavefunctions are indeed more localized at the domain walls (c.f. Figs.~\hyperref[mb50]{4(c-f)}).
The state shown in Fig. \hyperref[mb10]{4(c)} is the one originated from the chiral channels at the zigzag domain walls. Indeed, the current flows along a zigzag path, strictly following the geometry of the domain walls. And as a result of this current deformation, the local spin directions also exhibit a spatial texture (c.f. Fig.~\hyperref[mb50]{4(c)}).

Figs.~\hyperref[mb50]{4(e)} and \hyperref[mb50]{(f)} show the states with wavevectors in the x direction. The dispersion along this direction results from hopping between the domain walls. The zigzag domain walls largely facilitate the propagation in the x direction. These two states are constructed respectively with the chiral states on segments of the zigzag domain walls with the positive and negative projection of velocity in the x direction. Additionally, there is also some distribution of LDOS at the straight domain as an intermediate to couple the zigzag segments. The small local current flows along the straight domain walls compensate the y component of those at the zigzag segments, and the average current flow is in the x direction only.

\subsection{SK magnetization texture}

\begin{figure}[htbp]
    \centering
    \includegraphics[width=\linewidth]{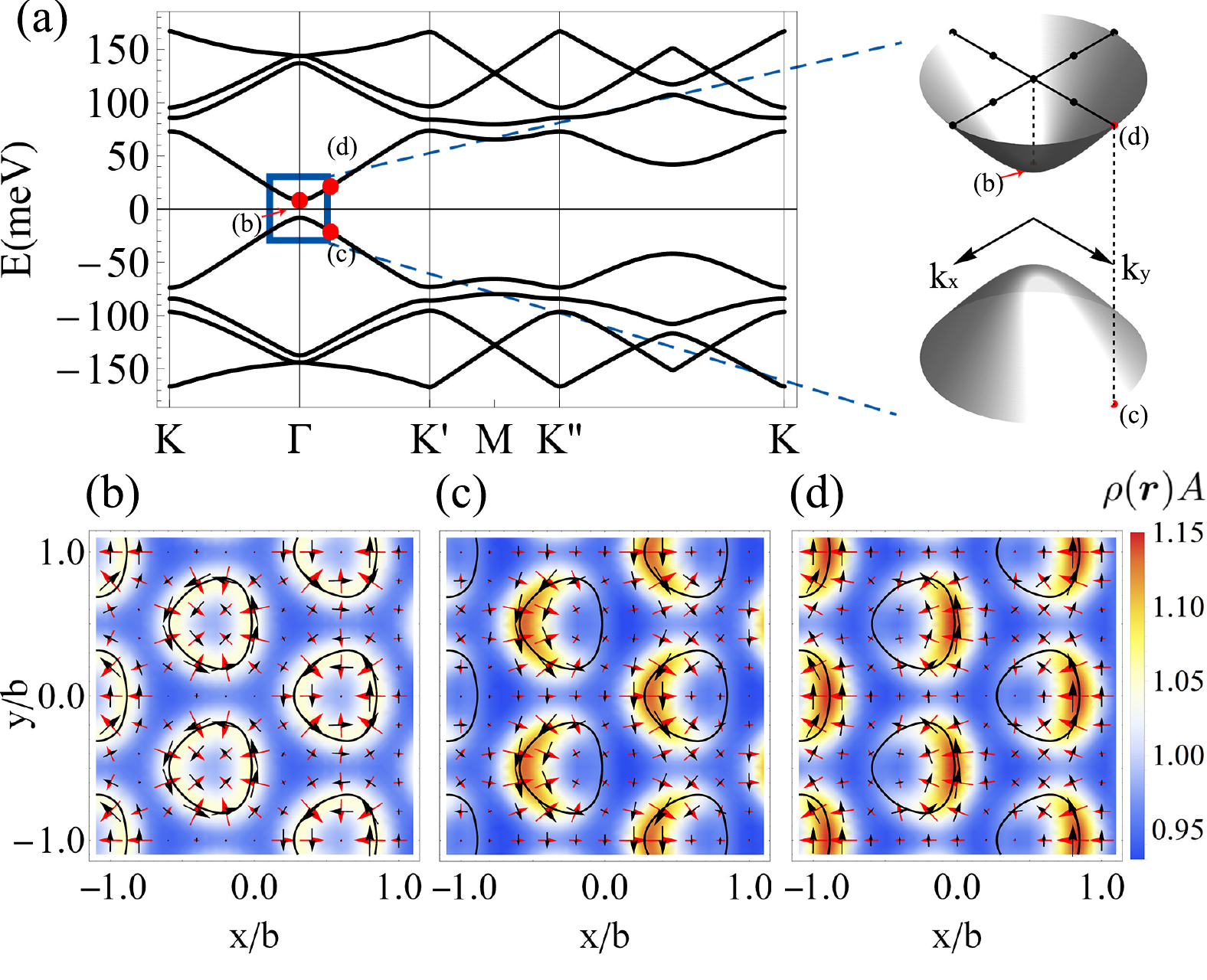}
    \caption{(a) The moir\'e minibands for a SK magnetization texture of moir\'e periodicity \(b=10\)nm. The right panel shows a full dispersion around the Dirac point. (b-d) LDOS $\rho(\bm{r})$ (background), electric current density \(\bm j(\bm{r})\) (black arrows) and in-plane components of local spin expectation \(\bm s(\bm{r})\) (red arrows) with wave vector \(k=0\) and \(1/b\) in y direction. The corresponding wave vector and band index are indicated by the red dots in (a). The wave function of conduction band can be derived from particle-hole symmetry, which is discussed in Sec.~\ref{summary}. The black lines are the domain walls. The parameters used are \(\hbar v = 2\text{eV}\cdot\text{\AA}\) and \(g=25\mathrm{meV}\).}\label{sk10}
\end{figure}

\begin{figure*}[htbp]
    \centering
    \includegraphics[width=0.85\linewidth]{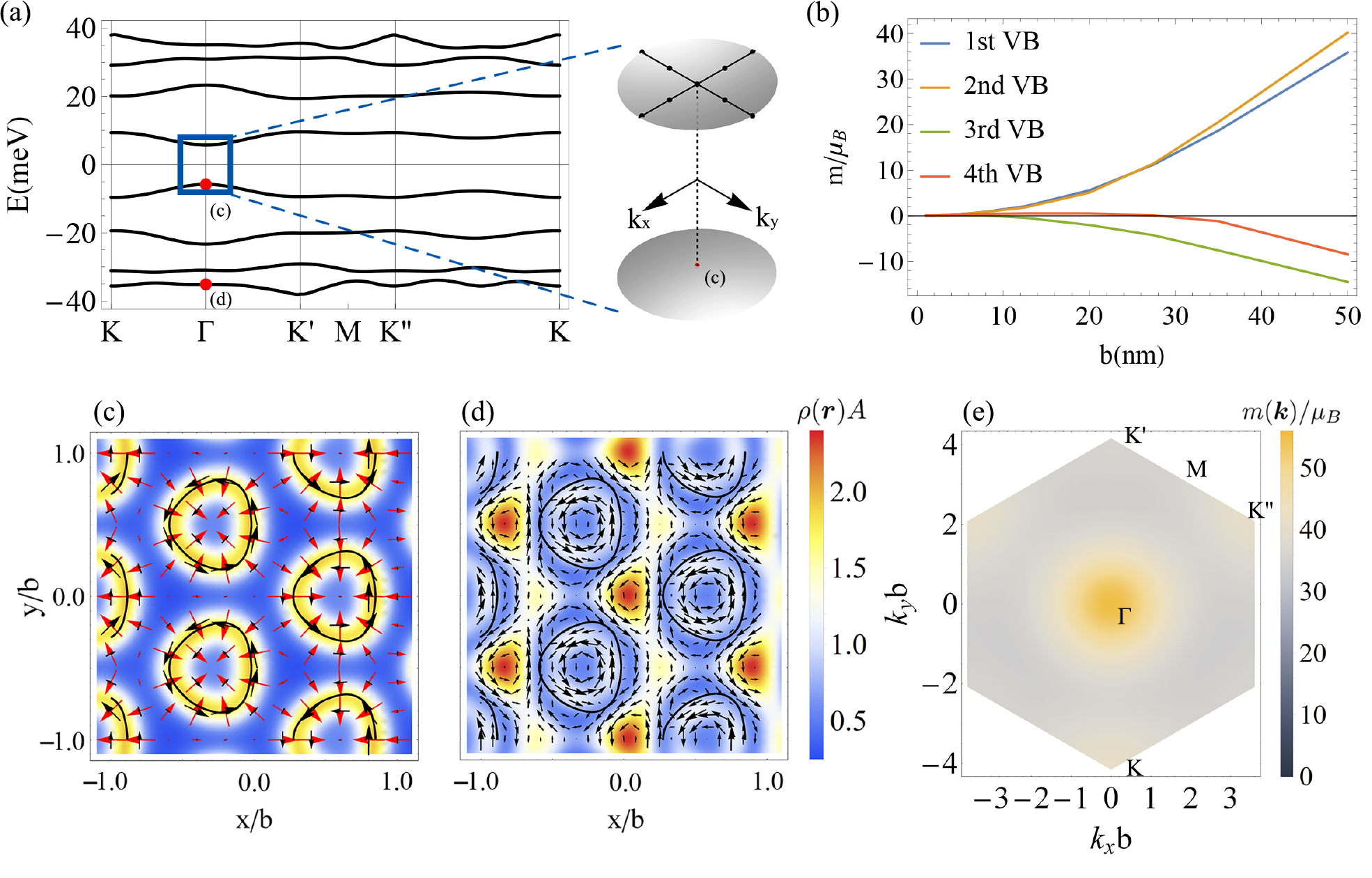}
    \caption{(a) The moir\'e minibands for a SK magnetization texture of moir\'e periodicity \(b=50\)nm. The right panel shows a full dispersion around the Dirac point. (b) The orbital angular momentum per mori\'e supercell for the four topmost valence bands as a function of moir\'e periodicity. (c-d) LDOS $\rho(\bm{r})$ (background), electric current density \(\bm j(\bm{r})\) (black arrows) and in-plane components of local spin expectation \(\bm s(\bm{r})\) (red arrows) for SK phase with wave vector \(k=0\). The corresponding wave vector and band index are indicated by the red dots in (a). The black lines are the domain walls. (e) The distribution of orbital magnetic momentum of the first valence band in the mini-Brillouin zone with moir\'e periodicity \(b=50\)nm. The parameters used are \(\hbar v = 2\text{eV}\cdot\text{\AA}\) and \(g=25\mathrm{meV}\).}\label{sk50}
\end{figure*}

The domain walls in SK magnetization texture are connected into closed loops and have $C_3$ rotation symmetry. Figs.~\hyperref[sk10]{5(a)} and \hyperref[sk50]{6(a)} show the moir\'e minibands in a SK magnetization texture with periodicity of 10 and 50 nm respectively. Compared with the MB case, one can see that the low-energy bands feature a gapped Dirac cone instead centered at the $\mathrm{\Gamma}$ point. For states right at the $\mathrm{\Gamma}$ point (c.f. Figs.~\hyperref[sk10]{5(b)} and \hyperref[sk50]{6(c)}), the LDOS is localized at the close loop domain walls with local current flowing counterclockwise around the enclosed domains of positive $M_z$ . The texture of local spin configuration in these states forms a vortex structure as a result of the spin-momentum locking. Away from the $\mathrm{\Gamma}$ point, a finite band velocity can be picked up due to the hopping between the chiral states at neighbouring domain walls. For states with finite wave vector (c.f.  Figs.~\hyperref[sk10]{5(c)} and \hyperref[sk10]{(d)}), the LDOS mainly distributes at the corresponding half of the domain wall loops, and the overall current does not average to zero. As the periodicity of the magnetization texture increases, the hopping gets suppressed and the miniband becomes flat with smaller band velocity (c.f. Figs. \hyperref[sk50]{6(a)}).

For these minibands originated from the chiral channels, the current circulation around the close loop domain walls would give rise to an orbital magnetization. For a given miniband Bloch state, the orbital magnetic moment can be defined: $\bm{m}\left(\bm{k}\right)=-i\frac{e}{2\hbar}\left\langle\nabla_{\bm{k}}u\middle|\times[H\left(\bm{k}\right)-\varepsilon(\bm{k})]\middle|\nabla_{\bm{k}}u\right\rangle$, where $\left|u\right\rangle$ is periodic part of the miniband Bloch function~\cite{chang1996berry}.
Fig.~\hyperref[sk50]{6(e)} plots the orbital magnetic moment in the top valence band, which is peaked at the $\Gamma$ point. Fig.~\hyperref[sk50]{6(b)} shows that the orbital magnetic moment integrated over the entire mini-Brillouin zone \(\bm{m}=\frac{A}{(2\pi)^2}\int_{BZ}\bm{m}\left(\bm{k}\right)d^2\bm{k}\) for the four valence bands, calculated at various periodicity of the magnetization texture. The first two bands show similar orbital magnetization per supercell, which grows quadratically with the periodicity $b$. These two bands are lying in the energy range $[-g, g]$, i.e. the maximal local gap created by the magnetization, and have their origins from the in-gap chiral states. The local current density forms current circulation along the close loop domain walls, and the magnetic moment from such current circular is proportional to the area of the loop, consistent with the $b^2$ dependence of the orbital magnetization from these two bands.
Fig.~\hyperref[sk50]{6(b)} also plots the orbital magnetization of the two valence bands away from the gap, which have smaller magnitude and reversed sign. Fig.~\hyperref[sk50]{6(d)} shows a wave function in one of these minibands. Similar to the MB case, such state does not have the chiral state origin, and the wavefunction is distributed in the regions between the close loop domain walls.

\section{DISCUSSION AND SUMMARY}\label{summary}

The qualitative difference between the minibands in MB and SK magnetization textures can be understood based on symmetry analysis. We note that the Hamiltonian \eqref{H} has particle-hole symmetry, \(\mathcal{S}\mathcal{H}_S(-\bm{k})\mathcal{S}^{-1}=-\mathcal{H}_S(\bm{k})\), where \(\mathcal{S}=\sigma_x\mathcal{K}\) with \(\mathcal{K}\) being the complex conjugate operator. As a result, the spectrums for both MB and SK spectrum should fulfill the particle-hole symmetry with a pair of states $\{E\left(\bm{k}\right),-E\left(-\bm{k}\right)\}$, which is consistent with our numerical results. Furthermore, the SK magnetization texture is $C_3$-rotation symmetric and resembles the circular symmetry for a small moir\'e pattern, hence the spectrum is almost isotropic in SK magnetization texture. In MB magnetization texture, $C_3$-rotation is broken, therefore the spectrum is highly anisotropic.

Different from the SK case, the low-energy spectrum in MB magnetization texture is gapless. This is due to the inversion antisymmetry of the MB magnetization texture, \(M_z(\bm{r})=-M_z(-\bm{r})\). Therefore, we have \(\mathcal{PHP}^{-1}=-\mathcal{H}\), where \(\mathcal{P}\) is space inversion. Around $\mathrm{\Gamma}$ point, the Hamiltonian \eqref{H} can be expressed as \(\mathcal{H}_{eff}(\bm k)=h_0(\bm{k})\mathbbm{1}+h_x(\bm k)\sigma_x+h_y(\bm k)\sigma_y+h_z(\bm k)\sigma_z\), in bases of the two lowest eigenstates $\{|\left.\bm{G}=0,\uparrow\right\rangle,|\left.\bm{G}=0,\downarrow\right\rangle\}$ at $\mathrm{\Gamma}$ point. The particle-hole symmetry requires,

\begin{equation}
    \begin{aligned}
        \mathcal{SH}(\bm k)\mathcal{S}^{-1} & =h_0(\bm{k})\mathbbm{1}+h_x(\bm k)\sigma_x+h_y(\bm k)\sigma_y-h_z(\bm k)\sigma_z \\
                                            & =-\mathcal{H}(-\bm k)
    \end{aligned}
\end{equation}

\noindent while the inversion asymmetry requires,

\begin{equation}
    \begin{aligned}
        \mathcal{PH}(\bm k)\mathcal{P}^{-1} & =h_0(\bm{k})\mathbbm{1}+h_x(\bm k)\sigma_x+h_y(\bm k)\sigma_y+h_z(\bm k)\sigma_z \\
                                            & =-\mathcal{H}(-\bm k)
    \end{aligned}
\end{equation}

\noindent These requirements lead to $h_z\left(\bm{k}\right)=0$, which implies a gapless spectrum at $\mathrm{\Gamma}$ point for the MB magnetization texture. Furthermore, we have $h_0\left(-\bm{k}\right)=-h_0\left(\bm{k}\right)$, $h_x\left(-\bm{k}\right)=-h_x\left(\bm{k}\right)$, and $h_y\left(-\bm{k}\right)=-h_y\left(\bm{k}\right)$, which means that these coefficients are odd orders of $\bm{k}$ and around $\mathrm{\Gamma}$ point are approximately linear. For SK magnetization texture this inversion antisymmetry is broken, $h_z\left(\bm{k}\right)$ is not required to be zero, and the spectrum is gaped. Another way to understand the gap here is the close loop geometry of the domain wall leads to energy quantization of the chiral channel.

\begin{figure}[htbp]
    \centering
    \includegraphics[width=\linewidth]{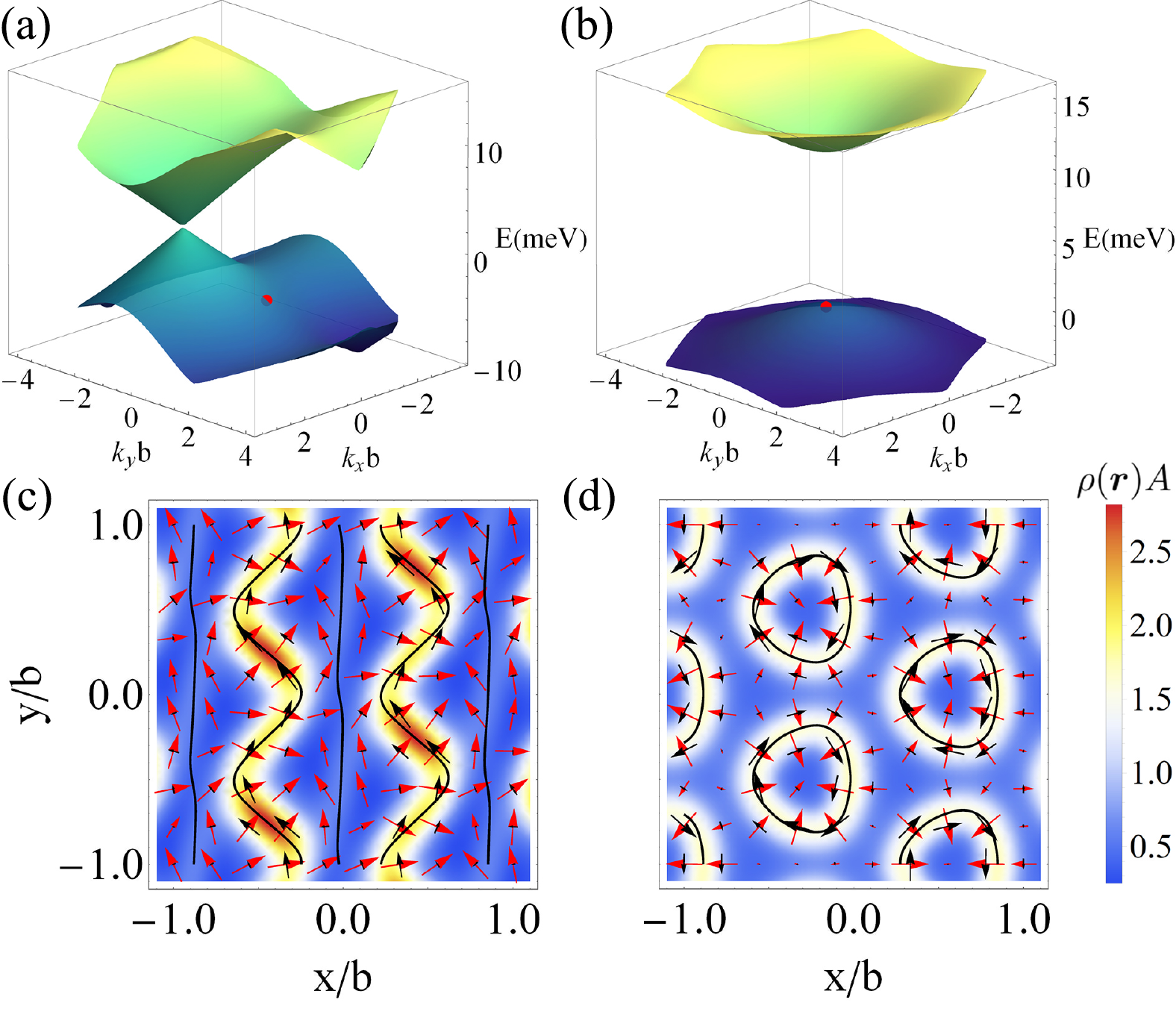}
    \caption{Minibands in the MB magnetization texture (a), and in the SK magnetization texture (b), when the in-plane magnetization is considered. (c) and (d) LDOS $\rho(\bm{r})$ (background), electric current density \(\bm j(\bm{r})\) (black arrows) and in-plane components of local spin expectation \(\bm s(\bm{r})\) (red arrows) for two representative Bloch states in (a) and (b) respectively (denoted by red dots). The black lines are the domain walls. The parameters used are \(g'=10\)meV and \(b=50\)nm.}
    \label{extended}
\end{figure}

Finally, we discuss the effect of the in-plane component of the magnetization texture on our results. Near the domain walls, the magnetization also has sizable in-plane components. To consider this effect, we extend the Hamiltonian \eqref{H} by adding the in-plane Zeeman term \(\mathcal{H}'=g' (M_x(\bm{r}) \sigma_x +M_y(\bm{r}) \sigma_y)\), where $M_x(\bm{r})$ and $M_y(\bm{r})$ are the in-plane magnetization. From Fig.~\hyperref[extended]{7}, one can see that adding this term simply shifts the gapless Dirac point in momentum space for the MB magnetization texture, while the effect on the wavefunction is not noticeable. For the SK magnetization texture, there is no noticeable effect on either the band dispersion or the wavefunction. This is because the in-plane magnetization of SK texture averages to zero with its $C_3$ rotational symmetry, while for the MB texture, there is a finite in-plane magnetization averaged over the supercell which displaces the Dirac point. These results can be understood in terms of the spin-orbit coupling induced by the magnetization texture. When the in-plane component is considered, the winding of the magnetization in the moir\'e pattern generates a non-Abelian field that yields synthetic spin-orbit coupling \cite{zhou2019tunable,narayan2019electrical,desjardins2019synthetic}, which would shift the Dirac point in the minibands.

In summary, we have studied the topological chiral states formed at the magnetic domain wall of the surface of a magnetic topological insulator. In the presence of a periodic magnetization texture that can arise by the formation of a long-period moir\'e pattern, the chiral domain wall states couple into a network and form  minibands. The miniband Bloch states feature a spatial texture of spin and local current density,
as clear manifestation of the spin-momentum locked chiral channels. The miniband features a gapless Dirac cone for the MB magnetization texture, while a gaped one for the SK magnetization texture, which is related to the symmetry of the magnetization texture. For SK magnetization texture, the current circulation carried by the chiral channel at the close loop domain walls leads to orbital magnetization. These results not only reveal an interesting interplay between surface state and the magnetization texture but also suggest a new platform to engineer spin and current flows through the manipulation of magnetization domains in spintronic devices.

\section{ACKNOWLEDGMENTS}

This work is supported by the Research Grants Council of Hong Kong (Grants No. HKU17306819 and No. C7036-17W), and the University of Hong Kong (Seed Funding for Strategic Interdisciplinary Research). Q.T. is supported by the National Natural Science Foundation of China (Grants No. 11904095) and the Fundamental Research Funds for the Central Universities from China.

\appendix
\section{}\label{A}

In this appendix, we study the magnetization reversal in the medium layer \ce{CrI_3} when stacked on an antiferromagnetic substrate \ce{MnPTe_3}. The first-principles calculations are implemented in the Vienna ab initio simulation package (VASP) with Perdew-Burke-Ernzerhof (PBE) functional \cite{kresse1996efficiency,bloch1994projector}. In order to account for strong electronic correlations for the Cr and Mn atoms, Hubbard on-site Coulomb parameters of 3eV and 4eV were respectively selected in the calculations. A vacuum layer with thickness of 20 \AA~is used to eliminate the interaction between the layers. The convergence criteria for energy and force are set as $10^{-5}$eV/\AA~and $10^{-2}$eV/\AA, the Monkhorst-Pack mesh is set to $7\times 7\times 1$, and a plane-wave cutoff energy of 500eV was used in the calculations. In addition, the interaction between layers for \ce{CrI_3}/\ce{MnPTe_3} heterostructures is described through vdW-D3 corrections \cite{stefan2010a}. We name the configuration when the Cr atoms of \ce{CrI_3} layer sit right on top of spin up (down) atoms of the \ce{MnPTe_3} layer as AB (BA) configuration. The calculated lattice constant and interlayer distance between the two planes containing magnetic atoms are 6.98 \AA~and 7.54 \AA~and are almost identical for AB and BA configuration. The calculated energy differences between the interlayer FM and AFM states are -6.06meV and 6.30meV for AB and BA configuration respectively. Therefore, in a long period moir\'e pattern, the magnetization tends to point at opposite direction at AB and BA locals. When \ce{MnBi_2Te_4} is stacked on this texture, because of the strong interlayer FM coupling between \ce{MnBi_2Te_4} and \ce{CrI_3}, the magnetization texture would also be formed in \ce{MnBi_2Te_4}.

\section{}\label{B}

The magnetization textures are calculated by adapting the parameters from the results in Ref. \cite{tong2018skyrmions}.
When a monolayer ferromagnet is stacked on an antiferromagnetic substrate, a moir\'e pattern forms at the interface between the monolayer ferromagnet and the antiferromagnetic substrate.
The effect of the antiferromagnetic substrate on the ferromagnet can be regarded as a spatially changing effective magnetic field, which prefers the formation of magnetic domains.
This effective field is determined by the interlayer magnetic interaction.
The steady-state magnetic configurations are determined by the competition between the effective field and intralayer magnetic interaction,
which can be solved from the coupled Landau-Lifshitz-Gilbert equations,
$\dfrac{d\bm{m}_i}{dt}=-\gamma\bm{m}_i\times\bm{H}_i^{eff}+\alpha\bm{m}_i\times\dfrac{d\bm{m}_i}{dt}$,
where $\bm{H}_i^{eff}=-\dfrac{\partial H}{\partial\bm{m}_i}$, and $\gamma$ and $\alpha$ are the gyro\-magnetic ratio and Gilbert damping coefficient respectively.
The Hamiltonian $H=-I\sum_{\langle i,j\rangle}{\bm{m}_i\bm{m}_j}-K\sum_{i}{\left(m_{z,i}\right)^2-\sum_{i}{\left[\bm{B}\left(R_i\right)\right]\cdot\bm{m}_i}}$, where $\bm{m}_i$ is magnetic moment at $i$-th lattice site and $\langle i,j\rangle$ covers all nearest neighboring sites.
$I>0$ is the intralayer exchange coupling and $K$ is the magnetic anisotropic energy.
The skyrmion and magnetic bubble configuration are both stable solutions for a long-period moir\'e pattern. The typical magnetization textures are shown in Figs. \hyperref[mass]{2(b)} and \hyperref[mass]{(c)}.
The parameters $K=0.014$I and the amplitude of effective field contributed from interlayer exchange and dipolar interaction $J_{ex}=5J_{dd}=0.13I$ are used.

%

\end{document}